\documentclass[twocolumn]{aastex631}

\usepackage[utf8]{inputenc}
\usepackage{amsmath,amssymb}
\usepackage{graphicx}
\usepackage{bm}

\begin{document}

\title{Detection of Lensed Gravitational Waves in the Millihertz Band Using Frequency-Domain Lensing Feature Extraction Network}

\author{Tianlong Wang}

\affiliation{School of Fundamental Physics and Mathematical Sciences, Hangzhou Institute for Advanced Study, University of Chinese Academy of Sciences (UCAS), Hangzhou 310024, China}
\affiliation{Center for Gravitational Wave Experiment, National Microgravity Laboratory, Institute of Mechanics, Chinese Academy of Sciences, Beijing 100190, China}
\affiliation{University of Chinese Academy of Sciences, Beijing 100049, China}

\author{Tianyu Zhao}
 \email{zhaotianyu@imech.ac.cn}
 \affiliation{Center for Gravitational Wave Experiment, National Microgravity Laboratory, Institute of Mechanics, Chinese Academy of Sciences, Beijing 100190, China}

 \author{Minghui Du}
 \affiliation{Center for Gravitational Wave Experiment, National Microgravity Laboratory, Institute of Mechanics, Chinese Academy of Sciences, Beijing 100190, China}

\author{Ziren Luo}
 \affiliation{Center for Gravitational Wave Experiment, National Microgravity Laboratory, Institute of Mechanics, Chinese Academy of Sciences, Beijing 100190, China}
 \affiliation{Key Laboratory of Gravitational Wave Precision Measurement of Zhejiang Province, Hangzhou Institute for Advanced Study, UCAS, Hangzhou 310024, China}
 \affiliation{Taiji Laboratory for Gravitational Wave Universe (Beijing/Hangzhou), UCAS, Beijing 100049, China}

 \author{Peng Dong}
 \email{dongpeng@ucas.ac.cn}
 \affiliation{School of Fundamental Physics and Mathematical Sciences, Hangzhou Institute for Advanced Study, University of Chinese Academy of Sciences (UCAS), Hangzhou 310024, China}
 \affiliation{Key Laboratory of Gravitational Wave Precision Measurement of Zhejiang Province, Hangzhou Institute for Advanced Study, UCAS, Hangzhou 310024, China}
 \affiliation{Taiji Laboratory for Gravitational Wave Universe (Beijing/Hangzhou), UCAS, Beijing 100049, China}

 \author{Peng Xu}
 \email{xupeng@imech.ac.cn}
 \affiliation{Center for Gravitational Wave Experiment, National Microgravity Laboratory, Institute of Mechanics, Chinese Academy of Sciences, Beijing 100190, China}
\affiliation{School of Fundamental Physics and Mathematical Sciences, Hangzhou Institute for Advanced Study, University of Chinese Academy of Sciences (UCAS), Hangzhou 310024, China}
 \affiliation{Key Laboratory of Gravitational Wave Precision Measurement of Zhejiang Province, Hangzhou Institute for Advanced Study, UCAS, Hangzhou 310024, China}
 \affiliation{Taiji Laboratory for Gravitational Wave Universe (Beijing/Hangzhou), UCAS, Beijing 100049, China}
 \affiliation{Lanzhou Center of Theoretical Physics, Lanzhou University, Lanzhou 730000, China}



\begin{abstract}

The space-based gravitational wave (GW) detectors are expected to observe lensed GW events, offering new opportunities for cosmology and fundamental physics.
Across the millihertz band, lensing effects transition from the wave-optics regime at lower frequencies to the geometric-optics approximation at higher frequencies.
Although traditional GW identification methods, such as matched filtering, are well established and effective, the intense computational resources required motivate the search for more efficient alternatives to accelerate candidate event screening. 
To address this bottleneck, we introduce a Dual-Channel Lensing feature extraction eXtended Long Short-Term Memory Network (DCL-xLSTM). 
Unlike conventional recurrent architectures, DCL-xLSTM uses a matrix-valued memory structure and a memory-mixing mechanism to effectively capture amplitude patterns that span the entire millihertz frequency band. 
Trained on data generated by Point Mass (PM) and Singular Isothermal Sphere (SIS) models accounting for the transition from wave-optics to geometric-optics, the proposed method achieves an area under the curve (AUC) exceeding 0.99, maintaining a true positive rate (TPR) above $98\%$ at a false positive rate (FPR) below $1\%$.
The network is robust against variations in signal-to-noise ratio, lens type, and lens mass, establishing its viability as a high-efficiency tool for future space-based GW detection.

\end{abstract}


\section{Introduction}

The first detection of the GW150914 event initiated the era of gravitational wave (GW) astronomy \cite{Abbott2016_GW150914}. Since then, the ground-based detector network has expanded the catalog to 200 confirmed GW signals \cite{Abbott2023_GWTC3,GWTC-4.0-1,GWTC-4.0-2,GWTC-4.0-3}, enabling unprecedented tests of general relativity in the strong-field regime \cite{Abbott2021_GR} and providing novel insights into astrophysical populations and merger rates of compact objects \cite{Abbott2023_Pop}.
Gravitational lensing has been verified by electromagnetic (EM) observations for decades and has led to several groundbreaking findings in astrophysics \cite{lin2025,Liu_2019,Kim2021,Li2025,Janquart2023,Savastano2024,Liao_2017,Liao_2022,cremonese2021}.Building on these EM successes, the lensing of GWs has emerged as a highly anticipated phenomenon, expected to occur across a diverse range of astrophysical sources, from compact binary coalescences to core-collapse supernovae \cite{ramesh2022gravitational}.

The next frontier in GW observation lies in the millihertz frequency band, which will be accessible to future space-borne interferometers such as LISA \cite{LISA2017}, Taiji \cite{Taiji2017,ren_taiji_2023,du_towards_2025}, and TianQin \cite{TianQin2016}.These observatories are expected to detect the mergers of massive black hole binaries (MBHBs) in high cosmological redshifts \cite{Cusin2021,Takahashi2003, Savastano2024,cremonese2021}. At these cosmological distances, the probability of strong gravitational lensing becomes significant, rendering the detection of lensed events not merely a possibility, but an expectation \cite{Savastano2024,_al_kan_2023}.
A key feature of GW lensing in the millihertz band is the critical role of wave-optics \cite{Villarrubia2024, Savastano2024}. The characteristic GW wavelength can be comparable to or larger than the Schwarzschild radius of the lens for lens masses $M_L \lesssim 10^8 M_\odot$, causing diffraction and interference effects to become significant \cite{Takahashi2003, Janquart2023, Liu2025,yuan2025}.

Early prospects for machine learning in GW astronomy were discussed by \cite{zhaotianyu}.For example,\cite{wang2024rapid,wang2025search,ramesh2022gravitational} applied deep neural networks to search for exotic GW signals and identify binary black hole mergers.
Recent studies have shown that deep learning frameworks have emerged as a high-speed alternative to matched-filtering for identifying lensed GW signals \cite{chan2025detectability,chan2026identification}.
An increasing amount of work has explored machine learning applications in GW lensing, ranging from rapid parameter estimation to spectrogram-based classification.

In the context of parameter estimation,  \cite{arxiv2412_00566} introduced a Conditional Variational Autoencoder (CVAE) for microlensed binary black hole waveforms. Their model reduces inference runtime by up to five orders of magnitude compared to Bayesian methods like Bilby, and incorporating CVAE-generated priors further cuts Bilby's runtime by about 48\% without sacrificing accuracy. \cite{arxiv2511_07186} extended the neural posterior estimation framework DINGO to develop DINGO-lensing, which recovers lensing parameters with millisecond precision. DINGO-lensing reconstructs the full posterior distribution in seconds rather than weeks and, importantly, can identify diffraction-dominated signals from point-mass lenses—directly bridging to the wave-optics regime discussed above.  \cite{arxiv2511_08486} combined the accurate diffraction integral code with DINGO-based neural posterior estimation, demonstrating efficient parameter inference for microlensed GWs under a point-mass lens model. 

For the task of identifying lensed GW signals,  \cite{Kim2021} utilized a VGG-19 network \cite{vgg} in spectrograms, successfully identifying lensing-induced "beating patterns" \cite{Bulashenko2022,Hou_2020,Hou_2021} with different SNR and performing parameter estimation for lens properties. The method typically reframes the identification task as an image classification problem, analyzing 2D time-frequency representations such as spectrograms or Q-transforms \cite{Goyal2021, Magare2024, Li2025, Kim2021}. The SLICK pipeline, introduced by \cite{Magare2024}, further enhanced this by using a parallel network architecture to analyze both the Q-transform and the Sine-Gaussian maps, finding that combined input significantly reduces false positives.
More recently, architectures have evolved to Vision Transformers \cite{VIT} . \cite{Li2025} proposed the Squeeze-and-Excitation Multilayer Perceptron Data-efficient Image Transformer model, which classifies spectrogram pairs by explicitly modeling their morphological similarity \cite{Li2025}. Specialized models, such as the Wavelet Convolutional Detector, have also been developed to specifically target the diffraction patterns associated with microlensing by compact dark matter \cite{Liu2025}. These collective works demonstrate the rapid maturation of deep learning as a powerful and efficient tool for GW data analysis. 

Existing machine learning approaches for identifying lensed GW signals, such as those based on analyzing 2D time-frequency spectrograms\cite{Goyal2021,Kim2021}, introduce significant computational overhead during the data transformation process. However, the time-frequency resolution trade-off inherent in spectrogram generation can under-resolve the fine-scale, oscillatory interference characteristics. This can effectively smooth out subtle spectral modulations that are key signatures for precise lensing identification, posing a fundamental bottleneck for methods reliant on this representation\cite{Hou_2020,Bulashenko2022,arxiv2411_12453}. Recent work has demonstrated that processing 1D strain sequences directly using convolutional neural networks can successfully identify lensed signals while bypassing the resolution bottlenecks of 2D time-frequency maps \cite{arxiv2411_12453}. Building upon this insight, analyzing the raw frequency-domain amplitude spectrum via a sequence-based modeling approach offers a highly robust alternative to retain the full fidelity of lensing effects.
 
In this work, we present a deep learning framework for the identification of lensed GWs in the millihertz band.
First, we develop a classifier that encompasses the continuous transition from the wave-optics to the geometric-optics regime. By extending beyond asymptotic limits, our dataset and model are designed to accurately capture complex amplitude modulation to ensure physical fidelity across the diverse lens masses relevant to LISA.
Second, we adopt a direct sequence modeling approach that leverages the full resolution of the frequency-domain amplitude spectrum. Unlike 2D image-based methods, where fine-scale spectral features may be attenuated due to resolution constraints, our method analyzes whitened  Time-Delay Interferometry (TDI) Channel A and E strain, which allows for the preservation of high-frequency oscillatory modulations, providing a robust basis for model identification.
Third, we employ a \textbf{D}ual-\textbf{C}hannel \textbf{L}ensing feature extraction e\textbf{X}tended \textbf{L}ong \textbf{S}hort-\textbf{T}erm \textbf{M}emory Network (\textbf{DCL-xLSTM}). Compared with conventional LSTM, it introduces a matrix-based memory structure and a memory-mixing mechanism that allows for more retention of intricate details in long spectral sequences. This architecture enhances the ability of the model to handle long-term dependencies beyond the capabilities of standard LSTMs, while maintaining linear computational complexity.

The remainder of this article is organized as follows. 
Section~\ref{sec:method} details the methodology, including the physics of lensing, the waveform simulation pipeline, and the DCL-xLSTM network architecture. 
Section~\ref{sec:results} presents the classification performance, analyzing the robustness of different lens models, masses, and signal-to-noise ratios. 
Finally, Section~\ref{sec:conclusion} summarizes our findings and discusses their implications for future multi-messenger astronomy.
\begin{figure*}[htbp] 
    \centering
    \includegraphics[width=1.0\linewidth]{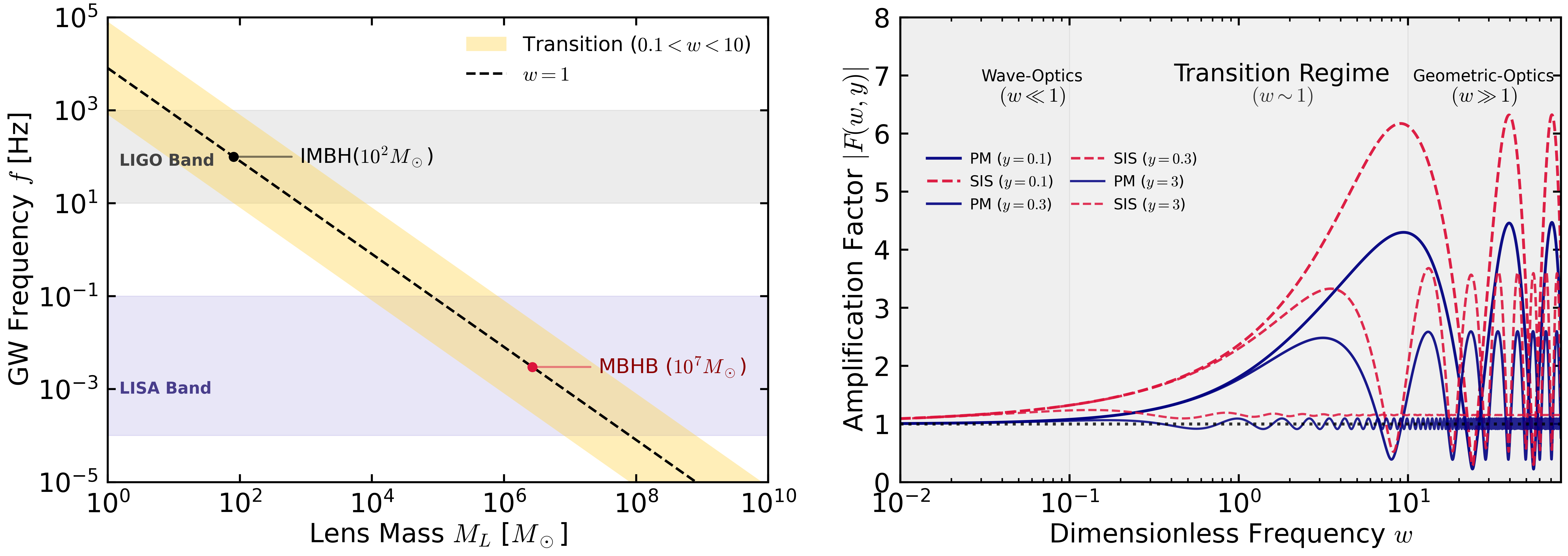}
    \caption{\label{fig:fw-all}
    \textbf{Overview of gravitational lensing regimes and signal amplification.}
    \textbf{Left:} The lens mass ($M_{L}$) versus GW frequency ($f$) parameter space. The dashed line ($w=1$) indicates the characteristic scale separating the wave-optics and geometric-optics limits, while the shaded orange band ($0.1 < w < 10$) highlights the approximate transition region. Sensitivity bands for LISA (blue) and LIGO (gray) are shown for reference.
    \textbf{Right:} The amplification factor $|F(w,y)|$ as a function of the dimensionless frequency $w$ for Point-Mass (PM, solid lines) and Singular Isothermal Sphere (SIS, dashed lines) lens models, evaluated at different impact parameters $y=0.1$, $0.3$, and $3$.The figure demonstrates that the detailed structure of the lensing amplification, including oscillatory interference patterns, depends on both $w$ and $y$.}
\end{figure*}

\section{Method}
\label{sec:method}

\subsection{Lens Models}


Gravitational lensing distorts GW signals through mechanisms largely determined by the interplay between the GW wavelength $\lambda_{\text{GW}}$ and the characteristic size of the lens (Schwarzschild radius $R_s$).
This relationship can be characterized by the dimensionless frequency parameter $w$ illustrated in Figure~\ref{fig:fw-all}.
Following the convention in the wave-optics literature \cite{Takahashi2003, Savastano2024}, we define $w$ in terms of the redshifted lens mass $M_{Lz} = M_L(1+z_L)$:
\begin{equation}
w = \frac{8\pi G M_{Lz} f}{c^3},
\end{equation}
as Figure~\ref{fig:fw-all} shows, the parameter divides the phenomena into two distinct regimes.
In the wave-optics regime ($w \lesssim 1$), diffraction effects dominate, causing amplitude oscillations and phase shifts without the formation of discrete geometric images.
In the geometric-optics regime ($w \gg 1$), the diffraction integral approximates a sum over discrete stationary points, manifesting as multiple images with magnifications and constant time delays.
However, the transition regime ($0.1 < w < 10$) bridges these extremes, corresponding to the scenario where the GW wavelength is comparable to the Schwarzschild radius of the lens.

As shown in the right panel, the transition regime produces the first prominent peak in the amplification factor $|F(w)|$ , which can be attributed to constructive interference.
This regime also marks the crossover from a smooth, non-oscillatory dominated behavior
at $w \lesssim 1$ to the oscillatory interference fringes characteristic of the geometric-optics limit at $w \gg 1$.
Moreover, for a fixed $w$, the amplification factor $|F(w,y)|$ decreases as the impact parameter $y$ increases, reflecting the weaker lensing effect when the source moves further from the lens axis. This $y$-dependence, together with the $w$-dependence shown in the figure, highlights that the full lensing phenomenology is governed by the two-dimensional parameter space $(w, y)$ rather than by $w$ alone.

At a fixed frequency $f$, the lensed waveforms $\widetilde{h}_{+,\times}^{L}(f)$ relate to the unlensed waveforms through the complex amplification factor $F(f)$:
\begin{equation}
\widetilde{h}_{+,\times}^{L}(f) = F(f)\,\widetilde{h}_{+,\times}(f).
\end{equation}

\begin{figure*}[htbp] 
    \centering
    \includegraphics[width=1\linewidth]{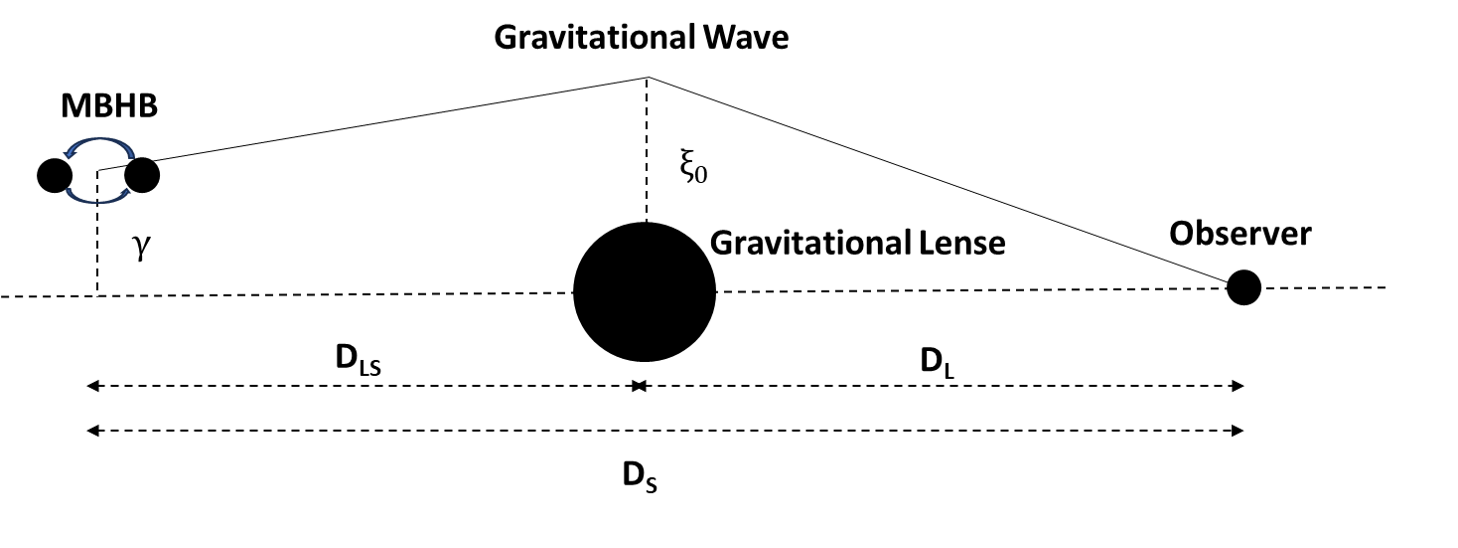}
    \caption{\textbf{A schematic diagram of gravitational lensing of GWs.} The signal from binary system is deflected by an intervening lens. Distances are shown: source-to-lens ($D_{LS}$), lens-to-observer ($D_L$). The impact parameter of the source relative to the lens axis is $\gamma$, and $\xi_0$ is the Einstein radius in the lens plane.}
    \label{fig:gw_lensing_diagram}
\end{figure*}

\subsubsection{Point Mass Lens}

 The point mass lens represents the simplest point case, characterized by a density profile $\rho(r) = M_L \delta^3(r)$ where $M_L$ denotes the lens mass, which is applicable to compact objects such as black holes. 
The amplification factor $F(w)$ is given by 
\cite{Takahashi2003,Matsunaga2006}:
\begin{equation}
\begin{aligned}
F(w) = &\exp\left[\frac{\pi w}{4} + \frac{i w}{2}\left(\ln\frac{w}{2} - 2\phi_m(y)\right)\right] \\
&\times \Gamma\left(1-\frac{i w}{2}\right) {}_1F_1\left(\frac{i w}{2}, 1, y^2\frac{i w}{2}\right),
\end{aligned}
\label{eq:pm_exact}
\end{equation}
where $\phi_{m}(y)=\dfrac{(x_{m}-y)^{2}}{2}-\ln x_{m}$ with $x_{m}=\dfrac{y+\sqrt{y^{2}+4}}{2}$. Here, $\Gamma(z)$ is the Euler gamma function and ${}_{1}F_{1}(a,b,z)$ is Kummer's confluent hypergeometric function.
The parameter $y$ represents the dimensionless position of the source, defined as $ y = \frac{\gamma D_{\mathrm{L}}}{\xi_{0}D_{\mathrm{S}}}, $
where $\gamma$ is the displacement of the source, $D_{\mathrm{L}}$ and $D_{\mathrm{S}}$ are the distances to the lens and to the source, respectively, and $\xi_{0} = \sqrt{(4GM_{\mathrm{L}}/c^{2})D_{\mathrm{LS}}D_{\mathrm{L}}/D_{\mathrm{S}}}$ is the Einstein radius of the lens, as illustrated in Figure~\ref{fig:gw_lensing_diagram}. The
analytical formula is able to give rise to the well-know approximations.

\subsubsection{Singular Isothermal Sphere Lens}

The SIS model, described by the density profile $\rho(r) = \sigma_v^2/(2\pi r^2)$ with $\sigma_v$ representing the velocity dispersion, which is a more complex representation suitable for galaxies or dark matter halos. The surface density is characterized as: $\Sigma(\xi) = \frac{\sigma_v^2}{2\xi}$
with the Einstein radius $\xi_0$ serving as normalization constant with $\xi_0 = 4\pi \sigma_v^2 \frac{D_L D_{LS}}{D_S}$.

The general solution for the amplification factor, valid across all physical regimes from wave-optics to geometric-optics, is given by the following diffraction integral \cite{Takahashi2003,Matsunaga2006}:

\begin{equation}
\begin{aligned}
F(f) &= -iwe^{iwy^2/2} \\
&\quad \times \int_0^\infty dx \, x \, J_0(wxy) \, \exp\left[ iw \left( \tfrac{1}{2}x^2 - x + \phi_m(y) \right) \right],
\end{aligned}
\label{eq:sis_exact}
\end{equation}
where $J_0$ is the zeroth-order Bessel function, $\phi_m(y) = y + 1/2$ and the lens mass is defined by:
\begin{equation}
M_{Lz}
=
(1+z_L)\,\frac{4\pi^2}{G\,c^2}\,
\sigma_v^{4}\,
\frac{D_L\,D_{LS}}{D_S},
\end{equation}
according to \cite{Hilker_2006,Hilker2004ThePO}, we take
$\sigma_v \simeq 20$--$40~{\rm km\,s^{-1}}$ for lenses of similar mass scale.
For the parameter $w>1$, the integrand becomes highly oscillatory. To ensure numerical stability and eliminate aliasing artifacts in the training data, we evaluate this integral using the Levin collocation method \cite{Levin1996}, which transforms the oscillatory quadrature into a non-oscillatory system of ordinary differential equations.

\subsection{Waveform Simulation}

We simulate lensed GWs originating from coalescing MBHB \cite{mbhb1,mbhb2}.
Our simulation pipeline proceeds as follows:
The frequency-domain waveforms $\tilde{h}_{+,\times}(f)$ are generated using the \textit{IMRPhenomD} Model \cite{IMRPhenomD1,IMRPhenomD2}. The source parameters are drawn from the distributions detailed in Table \ref{tab:full_parameter_space}.
We apply the complex amplification factor $F(f)$ directly to the source waveform: $\tilde{h}^{L}_{+,\times}(f) = F(f) \tilde{h}_{+,\times}(f)$. The step incorporates the frequency-dependent amplitude modulations derived from the PM or SIS models. And the lensed waveforms are projected onto the LISA constellation using the \textit{bbhx} \cite{bbhx1,bbhx2} software package, which computes the TDI observables $A, E$ and $T$ by accounting for the spacecraft's orbital motion and the frequency-dependent antenna response functions.
Finally, Gaussian noise colored by the LISA noise PSD $S_n(f)$ is added to the signals projected by the detector. The SNR is computed after lensing and projection to reflect the observed signal strength.

The parameter space is characterized by eight physical parameters:

\begin{equation}
\Theta = \{\eta, M, t_c, D_L(z_S), \theta_S, \phi_S, \iota, \psi\},
\end{equation}
where $\eta \equiv m_1 m_2/M^2$ is the symmetric mass ratio, $M \equiv m_1 + m_2$ denotes the total mass, and $D_L(z_S)$ represents the luminosity distance at the source redshift $z_S$. The angular parameters $(\theta_S, \phi_S)$ specify sky localization in the detector coordinates, while $\iota$ and $\psi$ determine the orbital inclination and polarization angles, respectively. Table~\ref{tab:full_parameter_space} summarizes the parameter ranges adopted for our simulations.

\begin{table}
\centering
\caption{Parameter space for the simulated lensed gravitational waves.}
\label{tab:full_parameter_space}

\begin{tabular}{lll}
\hline
Parameter & Range & Units \\
\hline

\multicolumn{3}{l}{\textit{1. GW Source Parameters}} \\
Source Mass ($M$) & $10^4$--$10^6$ & $M_\odot$ \\
Mass Ratio ($\eta$) & 0.06--0.25 & --- \\
Source Redshift ($z_S$) & 0.1--3.0 & --- \\
Sky Position $(\theta_S, \phi_S)$ & $[0,\pi]\times[0,2\pi]$ & rad \\
Inclination ($\iota$) & $0$--$\pi$ & rad \\
Polarization ($\psi$) & $0$--$\pi$ & rad \\
Coalescence Time ($t_c$) & $-3600$--$3600$ & s \\

\hline
\multicolumn{3}{l}{\textit{2. Lens Parameters}} \\
Lens Model & PM, SIS & --- \\
Lens Mass ($M_L$) & $10^6$--$10^8$ & $M_\odot$ \\
Lens Redshift ($z_L$) & 0.1--3.0 & --- \\
Impact Parameter ($y$) & 0.1--5.0 & --- \\

\hline
\multicolumn{3}{l}{\textit{3. Simulation Parameters}} \\
Waveform Model & IMRPhenomD & --- \\
Noise Model & LISA PSD & --- \\
SNR & 20--70 & --- \\

\hline
\end{tabular}
\end{table}

The detector projection combines GW polarizations through the frequency-domain response:

\begin{equation}
\tilde{h}^{A,E,T}(f) = \sum_{lm} \mathcal{T}^{A,E,T}(f,t_{lm}(f)) \tilde{h}_{lm}(f),
\end{equation}
where the time-frequency mapping follows from the stationary phase approximation:

\begin{equation}
t_{lm}(f) = t_{\text{ref}} - \frac{1}{2\pi}\frac{d\phi_{lm}(f)}{df}.
\end{equation}

The uncorrelated TDI channels are constructed as follows \cite{LISAaet}:

\begin{align}
A &= \frac{1}{\sqrt{2}}(Z-X), \\
E &= \frac{1}{\sqrt{6}}(X-2Y+Z), \\
T &= \frac{1}{\sqrt{3}}(X+Y+Z),
\end{align}
with $\mathcal{T}^{A,E,T}$ encoding both the antenna pattern and LISA's orbital motion. The implementation uses GPU-accelerated batch processing of harmonic modes and cubic spline interpolation for the response functions \cite{bbhx1,bbhx2}.


The noise characteristics of the frequency-domain are quantified by the one-sided power spectral density $S_n(f)$ \cite{LISA_PSD}.
\begin{align}
P_{\text{OMS}}(f) &= (15\,\text{pm})^2 \left[1 + \left(\frac{2\,\text{mHz}}{f}\right)^4\right] \left(\frac{2\pi f}{c}\right)^2 \, \text{Hz}^{-1}, \\
P_{\text{acc}}(f) &= (3\,\text{fm/s}^2)^2 \left[1 + \left(\frac{0.4\,\text{mHz}}{f}\right)^2\right] \nonumber \\
                 &\quad \times \left[1 + \left(\frac{f}{8\,\text{mHz}}\right)^4\right] \frac{1}{(2\pi f)^4}.
\end{align}
For the $A$ and $E$ TDI channels, PSDs can be derived as derived as \cite{LISA_PSD}

\begin{align}
S_{A0,E0}(f) &= 
\begin{aligned}[t]
&2\left[2\left(3 + 2\cos x + \cos 2x\right)P_{\text{acc}}(f) \right. \\
&\left. + (2 + \cos x)P_{\text{OMS}}(f)\right],
\end{aligned} \\
S_{A1,E1}(f) &= 4\sin^2 x\, S_{A0,E0}(f) ,\\
S_{A2,E2}(f) &= 4\sin^2 2x\, S_{A0,E0}(f),
\end{align}
where $x = 2\pi f L / c$, $L = 2.5\,\text{Gm}$ is the length of the LISA arms and $c$ is the speed of light.
Here, the subscripts $0$, $1$, and $2$ denote the 0th-, 1st-, and 2nd-generation TDI combinations (i.e., different TDI orders) for the $A$ and $E$ channels.
We adopt $S_{A2,E2}(f)$ in this work.

The SNR $\rho$ for a waveform $h(t)$ is computed using the inner product:

\begin{equation}
\rho = \sqrt{4\, \Re \int_{f_{\text{min}}}^{f_{\text{max}}} \frac{|\tilde{h}(f)|^2}{S_N(f)} df},
\end{equation}
where the integration interval $[f_{\text{min}}, f_{\text{max}}]$ corresponds to the sensitive band of LISA. Noise realizations are scaled to achieve target SNR while preserving the statistical properties of the LISA noise model.

\subsection{Dual-Channel Lensing Feature Extraction xLSTM}
\begin{figure*}[t] 
    \centering
    \includegraphics[width=1\linewidth]{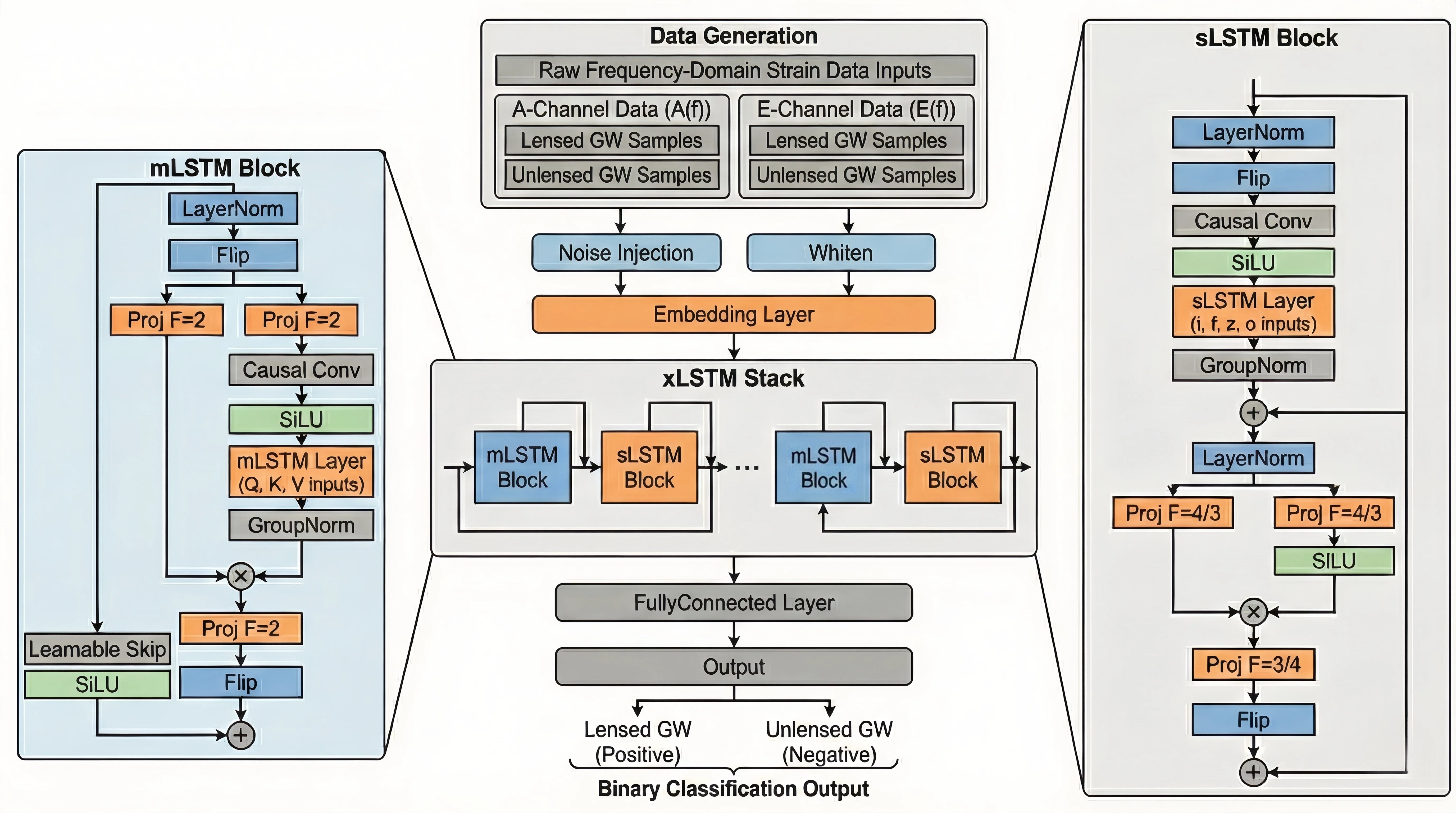}
    \caption{\textbf{The architecture of the DCL-xLSTM  for GW classification.} Frequency-domain strain amplitudes from the A and E TDI channels are preprocessed and sampled at 2048 points to form a dual-channel input sequence $\{\mathbf{x}_t\}$, where $\mathbf{x}_t = (|A(f_t)|, |E(f_t)|)$. The sequence is processed by a stack of mLSTM and sLSTM blocks, which extract long-range spectral features and cross-channel correlations characteristic of lensing. The final hidden representation is passed through a fully connected layer to produce a probability for lensed versus unlensed gravitational-wave events.}
    \label{fig:DCL-xLSTM}
\end{figure*}

Recurrent neural networks (RNNs) update a hidden state sequentially as
\begin{equation}
\mathbf{h}_t=\phi(\mathbf{W}\mathbf{x}_t+\mathbf{U}\mathbf{h}_{t-1}+\mathbf{b}),
\end{equation}
where \(\mathbf{x}_t\) is the input at step \(t\), \(\mathbf{h}_t\) is the hidden state, \(\phi(\cdot)\) is a nonlinear activation, and \(\mathbf{W},\mathbf{U},\mathbf{b}\) are trainable parameters.
Long Short-Term Memory (LSTM) networks \cite{LSTM} introduce a gated memory state \(\mathbf{c}_t\) and update
\begin{equation}
\begin{aligned}
\mathbf{f}_t &= \sigma(\mathbf{W}_f\mathbf{x}_t+\mathbf{U}_f\mathbf{h}_{t-1}+\mathbf{b}_f),\\
\mathbf{i}_t &= \sigma(\mathbf{W}_i\mathbf{x}_t+\mathbf{U}_i\mathbf{h}_{t-1}+\mathbf{b}_i),\\
\mathbf{o}_t &= \sigma(\mathbf{W}_o\mathbf{x}_t+\mathbf{U}_o\mathbf{h}_{t-1}+\mathbf{b}_o),\\
\tilde{\mathbf{c}}_t &= \tanh(\mathbf{W}_c\mathbf{x}_t+\mathbf{U}_c\mathbf{h}_{t-1}+\mathbf{b}_c),\\
\mathbf{c}_t &= \mathbf{f}_t\odot\mathbf{c}_{t-1}+\mathbf{i}_t\odot\tilde{\mathbf{c}}_t,\\
\mathbf{h}_t &= \mathbf{o}_t\odot\tanh(\mathbf{c}_t),
\end{aligned}
\end{equation}
where \(\sigma(\cdot)\) is the sigmoid function and \(\odot\) denotes element-wise multiplication.

We adopt the xLSTM architecture \cite{xLSTM}, which strengthens LSTM-style gating via exponential gates and employs two cell variants: sLSTM (vector-valued memory) and mLSTM (matrix-valued memory).
For both variants, gate pre-activations are parameterized by
\begin{equation}
\tilde{\mathbf{g}}_t=\mathbf{W}_g\mathbf{x}_t+\mathbf{U}_g\mathbf{h}_{t-1}+\mathbf{b}_g,
\qquad g\in\{i,f,o\},
\end{equation}
and exponential gating is applied element-wise, e.g.
\begin{equation}
\mathbf{i}_t=\exp(\tilde{\mathbf{i}}_t),\qquad
\mathbf{f}_t=\exp(\tilde{\mathbf{f}}_t),\qquad
\mathbf{o}_t=\sigma(\tilde{\mathbf{o}}_t).
\end{equation}
To stabilize exponential gates, xLSTM introduces a stabilizer state \(\mathbf{m}_t\) and uses stabilized gates \(\mathbf{i}'_t,\mathbf{f}'_t\) defined by
\begin{equation}
\begin{aligned}
\mathbf{m}_t &= \max(\log\mathbf{f}_t+\mathbf{m}_{t-1},\,\log\mathbf{i}_t),\\
\mathbf{i}'_t &= \exp(\log\mathbf{i}_t-\mathbf{m}_t),\\
\mathbf{f}'_t &= \exp(\log\mathbf{f}_t+\mathbf{m}_{t-1}-\mathbf{m}_t),
\end{aligned}
\end{equation}
where \(\max(\cdot,\cdot)\), \(\log(\cdot)\), and \(\exp(\cdot)\) are applied element-wise.

The sLSTM retains a vector-valued memory \(\mathbf{c}_t\in\mathbb{R}^{d_h}\) and introduces an explicit normalizer state \(\mathbf{n}_t\in\mathbb{R}^{d_h}\):
\begin{equation}
\begin{aligned}
\mathbf{c}_t &= \mathbf{f}'_t\odot\mathbf{c}_{t-1}+\mathbf{i}'_t\odot\tilde{\mathbf{c}}_t,\\
\mathbf{n}_t &= \mathbf{f}'_t\odot\mathbf{n}_{t-1}+\mathbf{i}'_t,\\
\mathbf{h}_t &= \mathbf{o}_t\odot\frac{\mathbf{c}_t}{\max(\mathbf{n}_t,\mathbf{1})},
\end{aligned}
\end{equation}
where \(\mathbf{1}\) is the all-ones vector and the division is element-wise.

The mLSTM replaces \(\mathbf{c}_t\) by a matrix memory \(\mathbf{C}_t\in\mathbb{R}^{d\times d}\).
Using a value vector \(\mathbf{v}_t\in\mathbb{R}^{d}\), the memory update takes an outer-product form
\begin{equation}
\mathbf{C}_t=\mathbf{f}'_t\mathbf{C}_{t-1}+\mathbf{i}'_t(\mathbf{v}_t\mathbf{v}_t^\top),
\qquad
\mathbf{v}_t=\mathbf{W}_v\mathbf{x}_t+\mathbf{U}_v\mathbf{h}_{t-1},
\end{equation}
where \((\cdot)^\top\) denotes transpose.

The hidden state is obtained by a normalized retrieval using a query vector \(\mathbf{q}_t\in\mathbb{R}^{d}\) and a normalization vector \(\mathbf{n}_t\in\mathbb{R}^{d}\):
\begin{equation}
\begin{aligned}
\mathbf{h}_t &= \mathbf{o}_t\odot
\frac{\mathbf{C}_t\mathbf{q}_t}{\max(|\mathbf{n}_t^\top\mathbf{q}_t|,\,1)},\\
\mathbf{q}_t &= \mathbf{W}_q\mathbf{x}_t,\\
\mathbf{n}_t &= \mathbf{W}_n\mathbf{x}_t+\mathbf{U}_n\mathbf{h}_{t-1},
\end{aligned}
\end{equation}
where \(|\cdot|\) denotes the absolute value.

The input consists of strain amplitude spectra from the A and E TDI channels.  
Each GW event is transformed into whitened frequency-domain amplitudes
\begin{align*}
\mathbf{A} = \{|A(f_1)|,\ldots,|A(f_{2048})|\},\\
\mathbf{E} = \{|E(f_1)|,\ldots,|E(f_{2048})|\},
\end{align*}
sampled at 2048 frequency points.  
The A and E channels respond differently to the same GW due to their distinct interferometric combinations and antenna pattern functions.  
To exploit this complementarity, we construct a two-dimensional feature vector at each frequency index $t$:
\[
\mathbf{x}_t = 
\big[\, |A(f_t)|,\ |E(f_t)| \,\big] \in \mathbb{R}^2,
\]
forming a dual-channel sequence $\{\mathbf{x}_t\}_{t=1}^{2048}$ that is fed into the first block.  
This representation enables the network to learn cross-channel correlations and identify coherent amplitude modulations characteristic of gravitational-wave lensing.

The final hidden state of the stack is passed through a fully connected layer to yield a lensing probability
\[
p = P(\text{lensed} \mid A(f), E(f)).
\]
A GW event is classified as lensed if $p > 0.5$,
and as unlensed otherwise.  
In an extended formulation, the network may additionally output channel-wise probabilities $p_A$ and $p_E$;  
a joint decision rule is then defined by
\[
p_{\mathrm{joint}} = p_A p_E,
\qquad
p_{\mathrm{joint}} > 0.5 \ \Rightarrow\ \text{lensed},
\]
reflecting the approximate statistical independence of noise in the A and E channels. The detailed architecture of the proposed network is illustrated in Figure~\ref{fig:DCL-xLSTM}, which presents the overall pipeline.

\begin{figure}[htbp]
    \centering
    \includegraphics[width=1\linewidth]{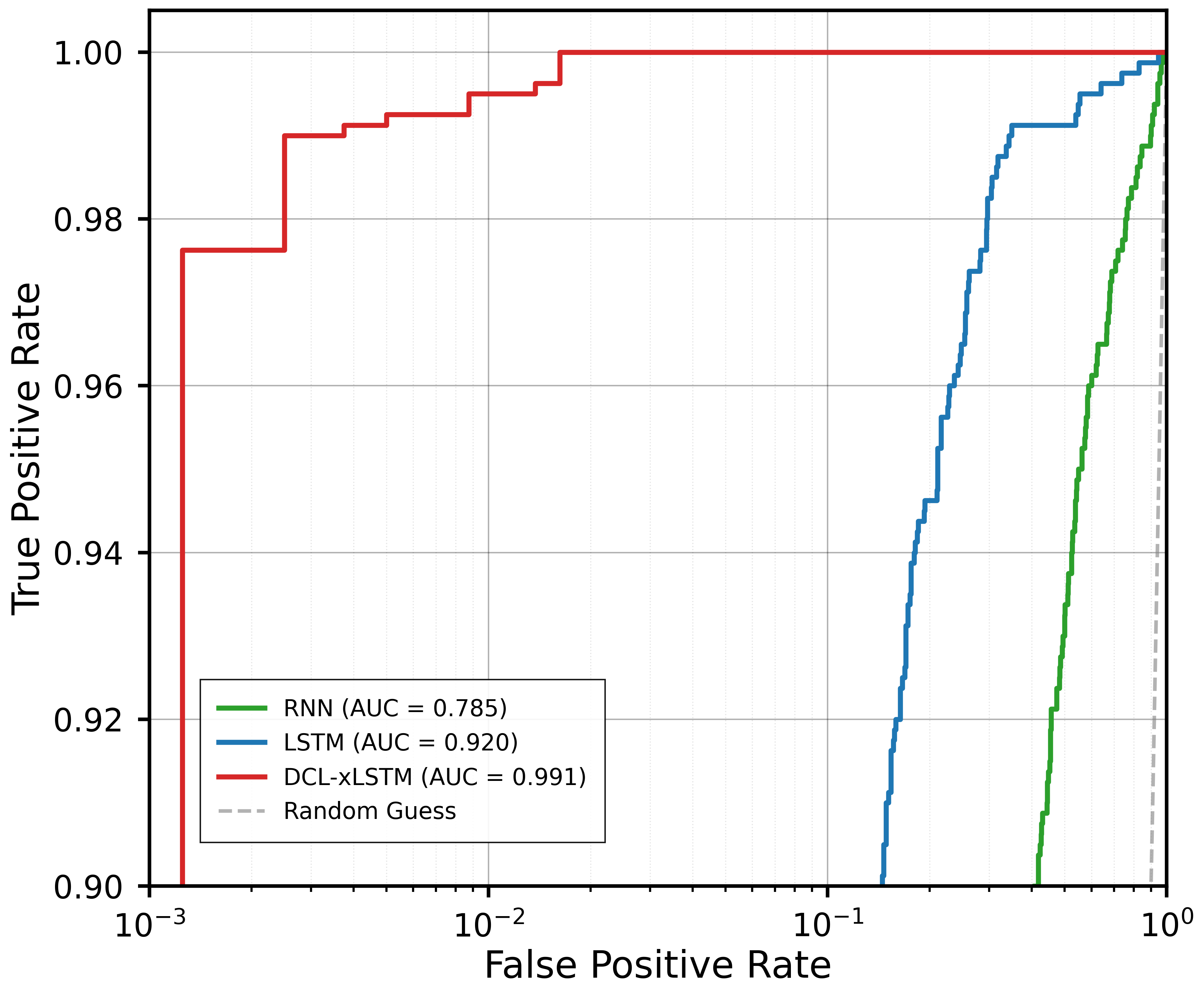}
    \caption{
    \textbf{Receiver operating characteristic (ROC) curves for the binary classification task on the combined dataset.} The DCL-xLSTM model (red solid line, AUC = 0.991) demonstrates  performance, significantly outperforming the LSTM (blue dashed line, AUC = 0.920) and the RNN (green dashed line, AUC = 0.785). The gray dashed line represents the random classifier baseline (AUC = 0.5). The x-axis (False Positive Rate) is plotted on a logarithmic scale to highlight performance at low FPRs, which is critical for rare event detection.
    }
    \label{fig:result1-overall}
\end{figure}

\begin{figure}[htbp] 
    \centering
    \includegraphics[width=1\linewidth]{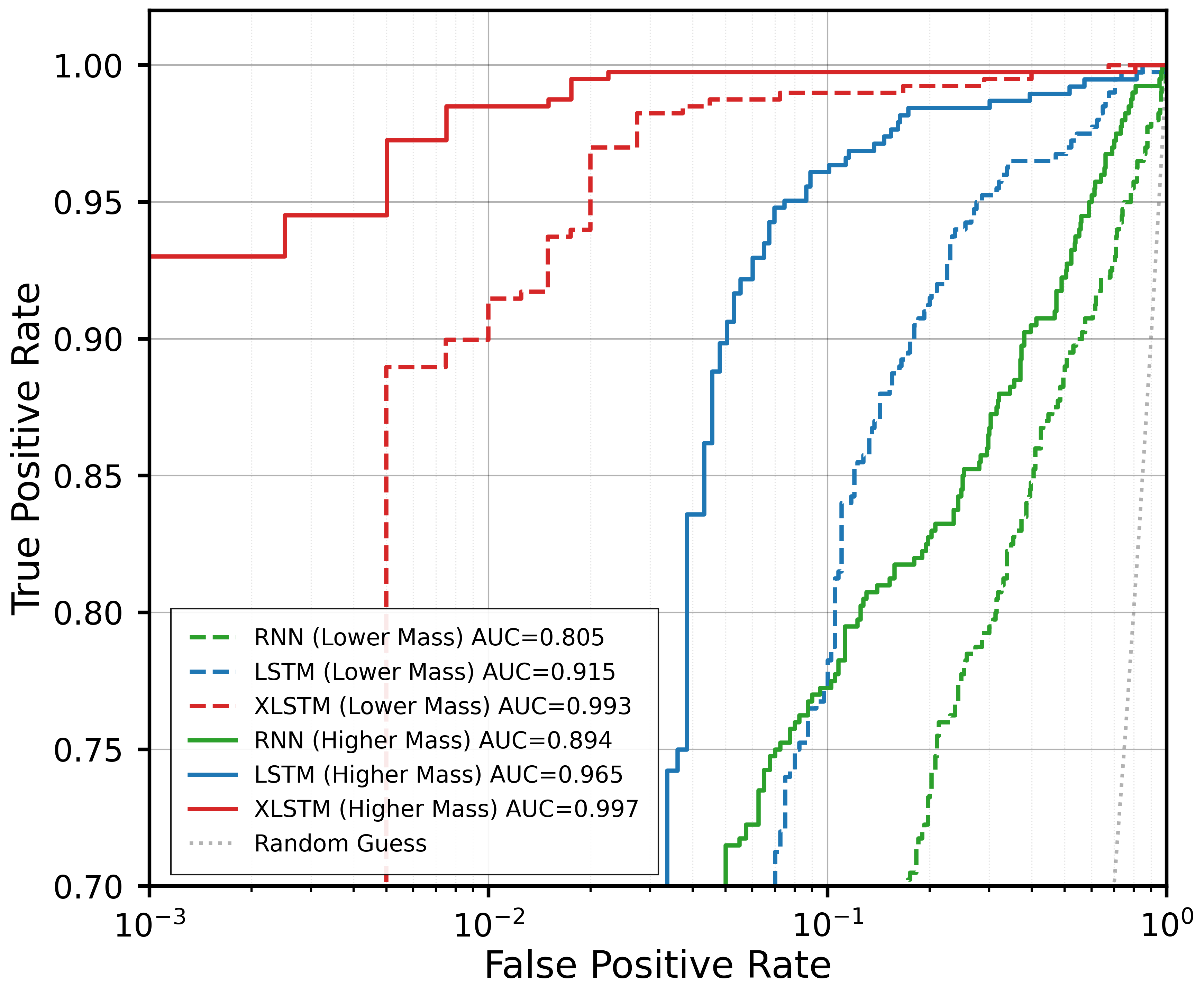}
    \caption{\label{fig:result2-mass}
    \textbf{Comparative ROC curves for GW signals classification.} The plot illustrates the performance of DCL-xLSTM (red), LSTM (blue), and RNN (green) models against the Higher Mass (solid lines) and Lower Mass (dashed lines) datasets. 
    While all models exhibit improved sensitivity for higher lens masses (solid curves), the DCL-xLSTM model displays stability, showing minimal performance degradation between mass regimes (AUC decreases only from 0.997 to 0.993). In contrast, the LSTM and RNN models show more significant performance gaps between the two datasets, highlighting the superior generalization capability of the DCL-xLSTM architecture.
    }
\end{figure}

\section{Results}
\label{sec:results}

To evaluate classification performance, we constructed a balanced dataset of 16,000 samples, consisting of equal numbers of lensed ($N=8,000$) and unlensed ($N=8,000$) waveforms.
We divided the samples into two distinct mass ranges to test the network's sensitivity across different conditions.
The \textit{High Mass} group ($M_L \in [10^7, 10^8] M_{\odot}$) represents a regime with strong wave-optics effects, where signal distortions are clearly visible.
In contrast, the \textit{Low Mass} group ($M_L \in [10^6, 10^7] M_{\odot}$) corresponds to the onset of diffraction, where the lensing features are subtle and the waveform deviations are lower.
This approach ensures that the model is tested against both clear and faint lensing signatures within the transition region.
The lensing effects were generated using two standard lens models: PM model \cite{Kim2021} and SIS model \cite{Villarrubia2024}.
To simulate realistic observation conditions, all signals were whitened and injected into Gaussian noise based on the LISA's noise model.
The optimal SNR was sampled from a uniform distribution of
30 to 70, allowing us to assess performance across a wide range of signal strengths.
The dataset was randomly divided into training (70\%), validation (15\%), and testing (15\%) sets, with strict separation to prevent data leakage.
We compared the proposed xLSTM-based classifier with the RNN and LSTM models.
The primary metric for performance is the area under the receiver operating characteristic curve (AUC), which measures the ability of the network to distinguish between classes independent of specific decision thresholds.

\subsection{General Classification Performance}
\label{subsec:performance overall}

The classification performance in the dataset is evaluated in Figure~\ref{fig:result1-overall}. The DCL-xLSTM model achieves near-perfect separability between lensed and unlensed classes, with an AUC of 0.991.
A practical advantage of the DCL-xLSTM classifier is its performance at low false positive rate (FPR). At an FPR of  \(10^{-3}\), it maintains a true positive rate (TPR) exceeding 0.98, which is critical for detecting rare lensing events with high confidence.

In contrast, standard LSTM (AUC = 0.920) and RNN (AUC = 0.785) exhibit significantly degraded performance in this low-FPR regime, with their TPR falling substantially below that of DCL-xLSTM.

The performance is consistent with the models' architectural capacity to capture the long-range, complex dependencies inherent in the representation of lensed waveforms. The better performance of DCL-xLSTM can be attributed to its matrix-valued memory and exponential gating mechanisms, which provide the necessary representational power to model the intricate correlations arising from the hybrid lensing physics. Based on the result, DCL-xLSTM thus serves as the optimal foundation for the detailed analyzes that follow.

\begin{figure*}[htb] 
    \centering
    \includegraphics[width=0.9\linewidth]{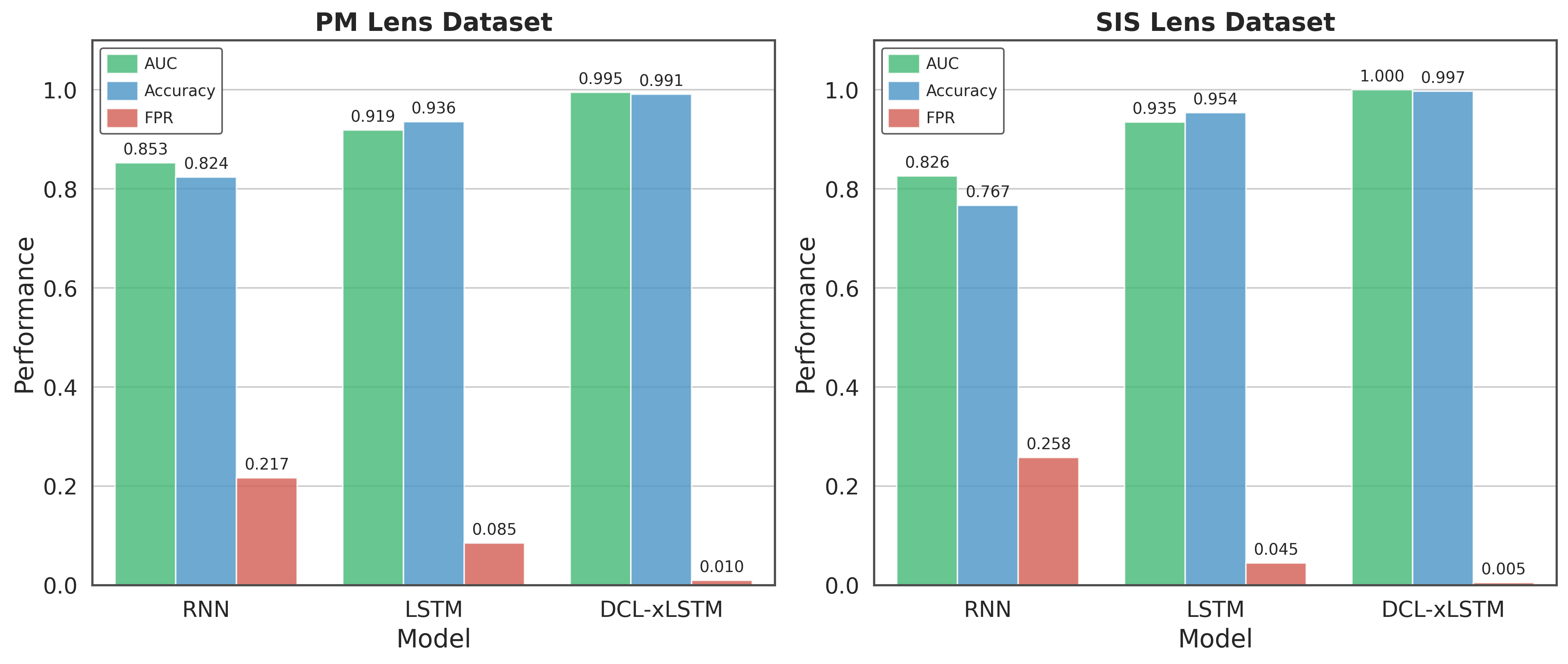}
    \caption{\textbf{Performance metrics (AUC, Accuracy, FPR) on different lens models.} \textbf{Left:} PM lenses. \textbf{Right:} SIS lenses. The DCL-xLSTM model achieves near-perfect AUC and accuracy  while maintaining a very low FPR (PM: 0.010, SIS: 0.005), outperforming LSTM and RNN across all metrics.}
    \label{fig:result3-lens}
\end{figure*}

\subsection{Robstness Across Lensing Regimes}

To assess the robustness of our approach under different diffractive conditions, we categorized the classification results according to lens mass.
The higher lens mass dataset ($M_L \in [10^7, 10^8] M_{\odot}$) highlights the regime in which wave-optics effects become pronounced, characterized by distinct modulations of amplitude.
In contrast, the lower lens mass dataset ($M_L \in [10^6, 10^7] M_{\odot}$) addresses the onset of diffraction. In this range, lensing signatures are inherently more subtle, resulting in waveform deviations that are less conspicuous than the higher mass counterparts.
The trends depicted in Figure~\ref{fig:result2-mass} illustrate how the strength of the diffractive features influences the classification efficacy.

In the higher lens mass regime, the pronounced spectral distortions induced by strong wave-optics effects provide clear discriminative features.
Consequently, all recurrent architectures operate effectively in this domain, with the RNN and LSTM achieving satisfactory sensitivity.
A more revealing divergence appears in the lower mass regime. As the lens mass decreases, the diffractive signatures become inherently more subtle, making them less distinguishable from the detector noise. Under these conditions, the performance of the baseline RNN and LSTM models begins to decrease.
In contrast, the DCL-xLSTM model retains stability and maintains a high AUC even when signal deviations are lower.
This resilience suggests that the matrix memory structure is particularly effective in capturing fine-grained transfer functions associated with the onset of diffraction, which may be overlooked by scalar-memory architectures.

To ensure the framework's applicability across different lens types, we extended our evaluation to include datasets generated with both PM and SIS lens profiles.
As illustrated in Figure~\ref{fig:result3-lens}, the DCL-xLSTM architecture exhibits remarkable consistency between these varying physical models.
Although the standard LSTM maintains competitive AUC scores, it struggles with false positives, exhibiting an FPR approximately $8.5\times$ (PM) and $9\times$ (SIS) higher than that of the DCL-xLSTM.
The baseline RNN faces greater challenges in this context, with AUC values dropping below 0.86 and the FPR exceeding 0.21, suggesting that simpler recurrent structures may fail to capture the lensed features from noise in diverse density profiles. The high stability and low false alarm rate of the DCL-xLSTM underscore its potential for GW analysis under varied astrophysical conditions.

\begin{figure*}[htb] 
    \centering
    \includegraphics[width=1\linewidth]{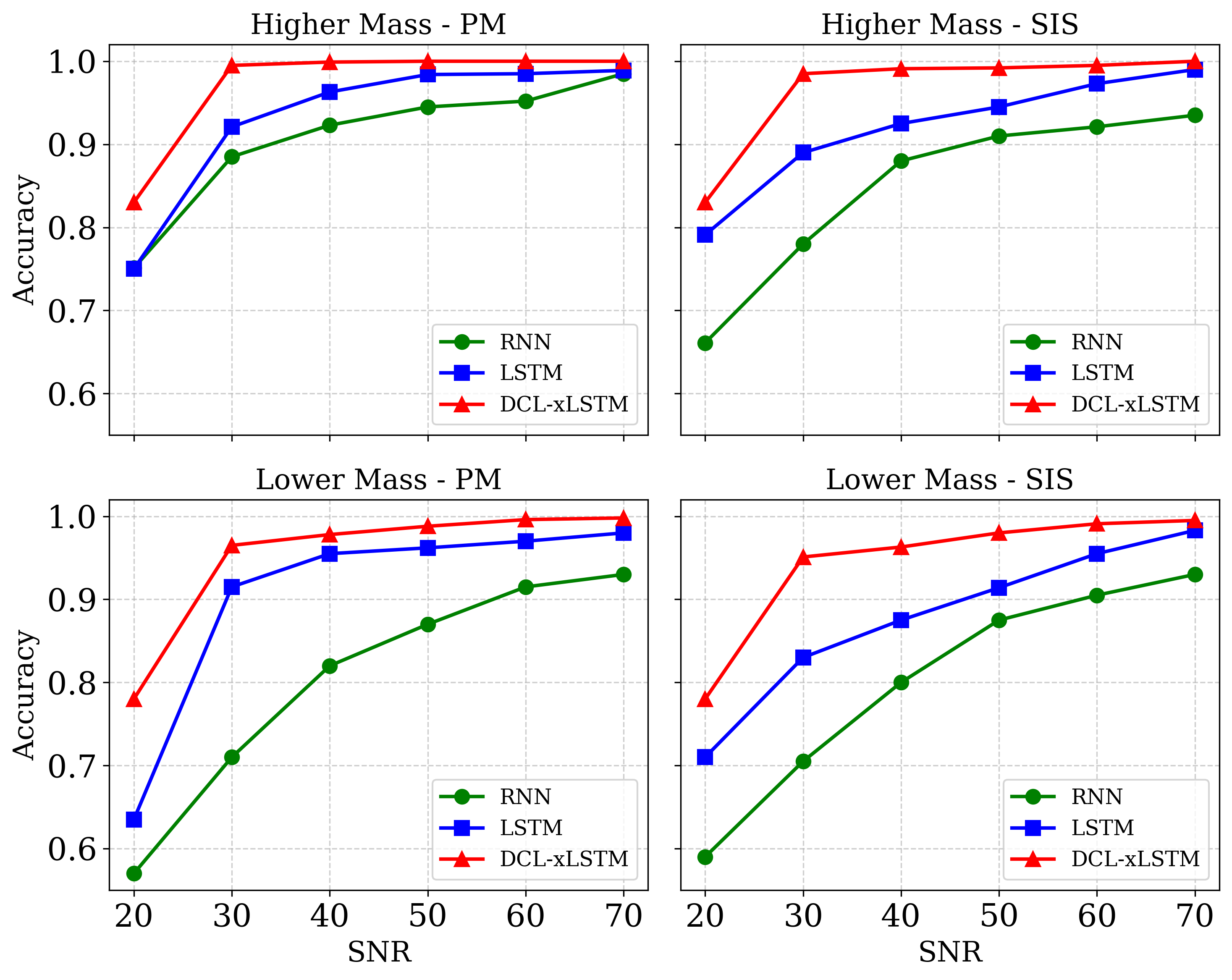}
    \caption{\textbf{Classification accuracy versus SNR for the RNN (green circles), LSTM (blue squares), and DCL-xLSTM (red triangles) models. } The four panels correspond to different combinations of source mass (higher/lower) and lens model (PM/SIS). The DCL-xLSTM model maintains the highest accuracy across all SNR levels and physical scenarios, with the performance advantage being most pronounced at low SNR.}
    \label{fig:snr}
\end{figure*}

\subsection{Detection Sensitivity Across SNR}

The sensitivity of a detection algorithm to varying noise conditions is important for practical applications. We evaluated the classification accuracy of the RNN, LSTM, and DCL-xLSTM models at SNR values of 30, 40, 50, 60, and 70. We also included SNR = 20 to assess performance at lower SNR and to analyze the trend of accuracy as the SNR increases.

The results, summarized in Figure~\ref{fig:snr}, show that the classification accuracy of three models improves as the SNR increases.
Across the two lens models, the accuracy trends with SNR are consistent and the model ranking is stable. When the lens masses are comparable, differences between lens models appear to be modest, which is plausible for a binary classification task where performance is driven mainly by how clearly the lensing signature emerges above the noise.
In the higher lens mass regime, all models achieve high accuracy (\(>0.90\)) at moderate SNR levels (\(\geq 50\)). The DCL-xLSTM model saturates this comparable performance level at a lower SNR (\(\sim 30\)) compared to LSTM and RNN.
In the lower lens mass regime, the lensing signature is a weaker spectral distortion, which makes its characteristic features inherently more subtle and less distinguishable from noise. The inherent difficulty leads to a more significant performance gap between models, especially at lower SNR (20-30).
Within this regime, DCL-xLSTM tends to retain a clear accuracy advantage over both LSTM and RNN, with the difference most pronounced at the lowest SNR. This pattern is consistent with improved robustness to noise and a stronger ability to capture distortion-related structure under unfavorable observational conditions.

\begin{figure*}[htb] 
    \centering
    \includegraphics[width=1\linewidth]{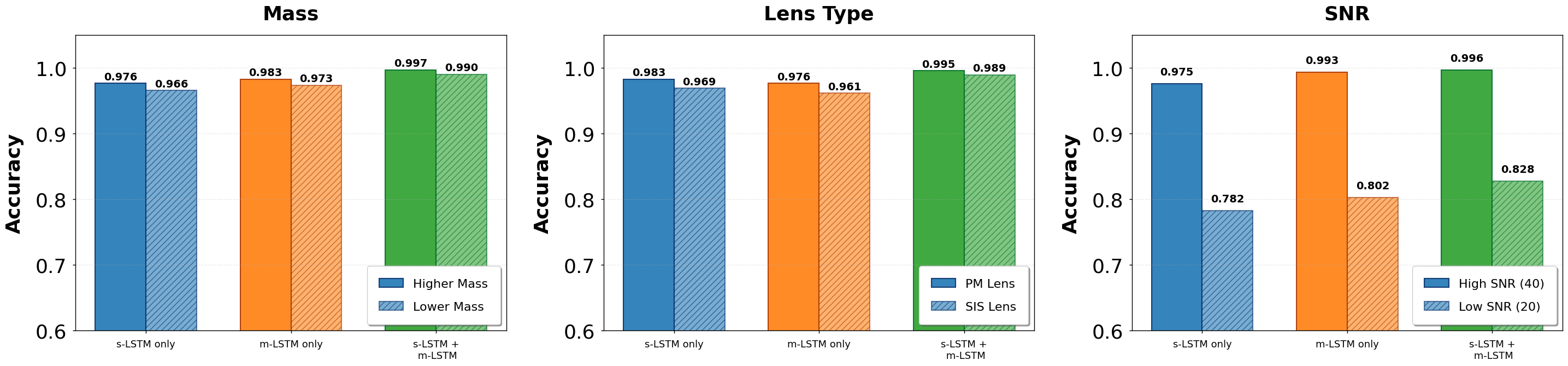}
    \caption{
    \textbf{
     Performance comparison of DCL-xLSTM architectural variants across different physical conditions.} The hybrid (s+m-LSTM) model leads the performance among all architectural variants, particularly in challenging low-SNR and complex lensing scenarios.
    }
    \label{fig:xlstm_variants}
\end{figure*}
\subsection{Ablation Study}

Given the consistently strong performance of DCL-xLSTM in the previous sections, we further examine the contributions of its core architectural components.
The xLSTM-based architecture integrates two types of memory blocks: sLSTM and mLSTM. To isolate their individual and synergistic effects, we conducted an ablation study comparing three variants: an sLSTM-only model, an mLSTM-only model, and the full hybrid model. The hybrid model, which was selected as the best-performing configuration after an extensive hyperparameter search, was used in all prior experiments.

The results of this ablation study are summarized in Figure~\ref{fig:xlstm_variants}. 
All three variants achieve high and comparable accuracy under low-SNR conditions and in wave-optics effects scenarios, indicating that the distinct lensing features can be learned by either architectural paradigm.
A significant performance gap emerges in the more challenging regimes, specifically under low-SNR conditions with less wave-optics effects. In these cases, the hybrid model consistently outperforms both single-component variants. For example, at an SNR of 20, the hybrid model achieves an accuracy approximately 4\% higher than the single-component variants. The result provides empirical evidence for functional complementarity between the sLSTM and mLSTM blocks when processing noisy, complex signals.

The sLSTM block excels at modeling fine-grained temporal dependencies through its scalar memory and enhanced gating mechanisms. The mLSTM block captures large-scale global patterns efficiently via its parallelizable matrix-valued memory. The superior performance of the hybrid model under diverse conditions demonstrates its ability to take advantage of these complementary strengths, validating its selection as the optimal architecture for the classification of lensed GW signals.

\section{Conclusion and Discussion}
\label{sec:conclusion}
We have developed and validated a deep learning framework for the identification of lensed GW signals in the millihertz band. The proposed DCL-xLSTM architecture consistently outperforms standard recurrent models (LSTM and RNN), achieving high classification performance in challenging scenarios involving mixed datasets. The DCL-xLSTM model maintains robust sensitivity across a range of lens masses and SNR, including regimes near the transition between wave-optics and geometric-optics, where classification tasks become more difficult. These results suggest that the architecture may serve as a useful tool for future space-based GW lensing studies.

While the proposed framework already demonstrates strong classification capabilities, there are still many directions that could further enhance its practical efficacy under more realistic conditions. In this work, we have adopted \textit{IMRPhenomD} waveforms for data generation. Moving forward, incorporating additional waveform templates, such as those including higher-order modes, spin precession, and EOB-based models, may help to broaden physical coverage and improve generalization.
Another aspect worth exploring is the inclusion of confusion noise, particularly the unresolved Galactic binaries expected to dominate the millihertz band. Taking this factor into account  during training could help the model better distinguish lensing-induced features from foreground structures.
Additionally, while current experiments assume stationary Gaussian noise, incorporating more realistic noise models that reflect mission operations may improve the framework performance and reliability. These directions represent incremental yet valuable extensions to build upon the current work.

In future work, we plan to explore two extensions to broaden the scope of the present study.
First, the framework can be extended beyond the single-plane, spherically symmetric lens assumption to incorporate more realistic lens configurations, including multi-plane lensing and composite lenses composed of multiple compact objects and extended mass distributions. Tools such as the GLoW package would support this expansion and allow the classifier to be tested with more scenarios. 
Second, the transverse motion of the lens and source is an attractive next step. Incorporating the dynamical effects into the training pipeline would broaden the set of measurable lensing observables and enable joint inference of lens parameters and effective transverse velocity.
These developments would improve the realism of the simulations and strengthen the detection of lensed signals in millihertz GW observations.

\bibliography{referencesAPJ} 
\end{document}